\documentclass[10pt,twocolumn,english,superscriptaddress,nofootinbib,amssymb,amsmath,prb,aps,floatfix]{revtex4-1}
\usepackage[latin9]{inputenc}
\usepackage{xcolor}
\usepackage{pdfcolmk}
\usepackage{units}
\usepackage{graphicx}
\PassOptionsToPackage{normalem}{ulem}
\usepackage{ulem}
\usepackage[english]{babel}
\usepackage{verbatim}
\usepackage{hyperref}

\makeatletter

\providecommand{\tabularnewline}{\\}
\newcommand{\mox}{Mo$_8$O$_{23}$ }

\makeatother

\usepackage{babel}
\begin{document}

\title{Time-resolved reflectivity and Raman studies of the interplay of
enigmatic orders in Mo$_{8}$O$_{23}$ }

\author{V. Nasretdinova}

\email{Venera.Nasretdinova@ijs.si}

\affiliation{Jozef Stefan Institute, Jamova 39, 1000 Ljubljana, Slovenia}

\affiliation{Center of Excellence on Nanoscience and Nanotechnology Nanocenter
(CENN Nanocenter), Jamova 39, 1000 Ljubljana, Slovenia}

\author{M. Borov\v{s}ak}

\affiliation{Jozef Stefan Institute, Jamova 39, 1000 Ljubljana, Slovenia}

\author{J. Mravlje}

\affiliation{Jozef Stefan Institute, Jamova 39, 1000 Ljubljana, Slovenia}

\author{P. \v{S}utar}

\affiliation{Jozef Stefan Institute, Jamova 39, 1000 Ljubljana, Slovenia}

\author{E. Goreshnik}

\affiliation{Jozef Stefan Institute, Jamova 39, 1000 Ljubljana, Slovenia}

\author{T. Mertelj}

\affiliation{Jozef Stefan Institute, Jamova 39, 1000 Ljubljana, Slovenia}

\affiliation{Center of Excellence on Nanoscience and Nanotechnology Nanocenter
(CENN Nanocenter), Jamova 39, 1000 Ljubljana, Slovenia}

\author{D. Mihailovic}

\affiliation{Jozef Stefan Institute, Jamova 39, 1000 Ljubljana, Slovenia}

\affiliation{Center of Excellence on Nanoscience and Nanotechnology Nanocenter
(CENN Nanocenter), Jamova 39, 1000 Ljubljana, Slovenia}

\affiliation{Faculty of Mathematics and Physics, University of Ljubljana, Jadranska
19, 1000 Ljubljana, Slovenia}

\date{\today}
\begin{abstract}
Monoclinic semi-metallic \mox belongs to a multifunctional series
of compounds showing multiple ordering phenomena that have not achieved
much attention till now. Previous X-rays studies of this compound
have revealed an incommensurate ordering transition at $T_{\mathrm{IC}}\sim350$\,K,
followed by a structural transition to commensurate order at $T_{\mathrm{IC-C}}=285$\,K.
In addition, an enigmatic resistance maximum is observed
at $T_{\mathrm{el}}\sim150$\,K, whose origin has
so far proved elusive. Aiming to disentangle these multiple orders
we use the polarized transient optical spectroscopy supplemented by
Raman spectroscopy to study the electronic relaxation dynamics and
lattice vibrational modes in \mox single crystals. Remarkably, both
the coherent vibrational mode response and single particle response
display extrema of damping/relaxation times close to $T_{\mathrm{el}}$
with the concurrent appearance of new coherent vibrational modes and
a characteristic polarization asymmetry which saturates below $T_{\mathrm{el}}$.
The single-particle relaxation data analysis shows the appearance
of a temperature-independent gap in the electronic excitation spectrum
below $T_{\mathrm{IC}}$ and additional temperature-dependent gap opening near $T_{\mathrm{el}}$.
Concurrently, a low frequency vibrational mode shows anomalous softening
around $T_{\mathrm{m}}\sim200$\,K, far below {\normalsize{}{}$T_{\mathrm{IC-C}}$}
and {\normalsize{}{}$T_{\mathrm{IC}}$}. The observations are interpreted
in terms of the appearance of a hidden gapped state below $T_{\mathrm{el}}$
that has so far eluded detection by structural analyses. 
\end{abstract}
\maketitle

\section{Introduction}

Low-dimensional molybdenum oxide \mox belongs to the family of MoO$_{3-x}$
suboxides (Magneli phases), which have been studied as perspective
charge storage or battery materials \cite{Lithiumo8o23,Lithiumbatt,MoO3-xnature}.
More recently, the ideas of memory devices based on electrical switching/topotactic
transitions in molybdenum suboxides have been discussed \cite{patents,MiloshAFM}.
In spite of increased recent interest in their functional properties,
the phase diagrams, transient and even equilibrium optical properties
have not been systematically studied for many molybdenum suboxides.
\begin{figure}[ht]
\includegraphics[clip,width=1\columnwidth]{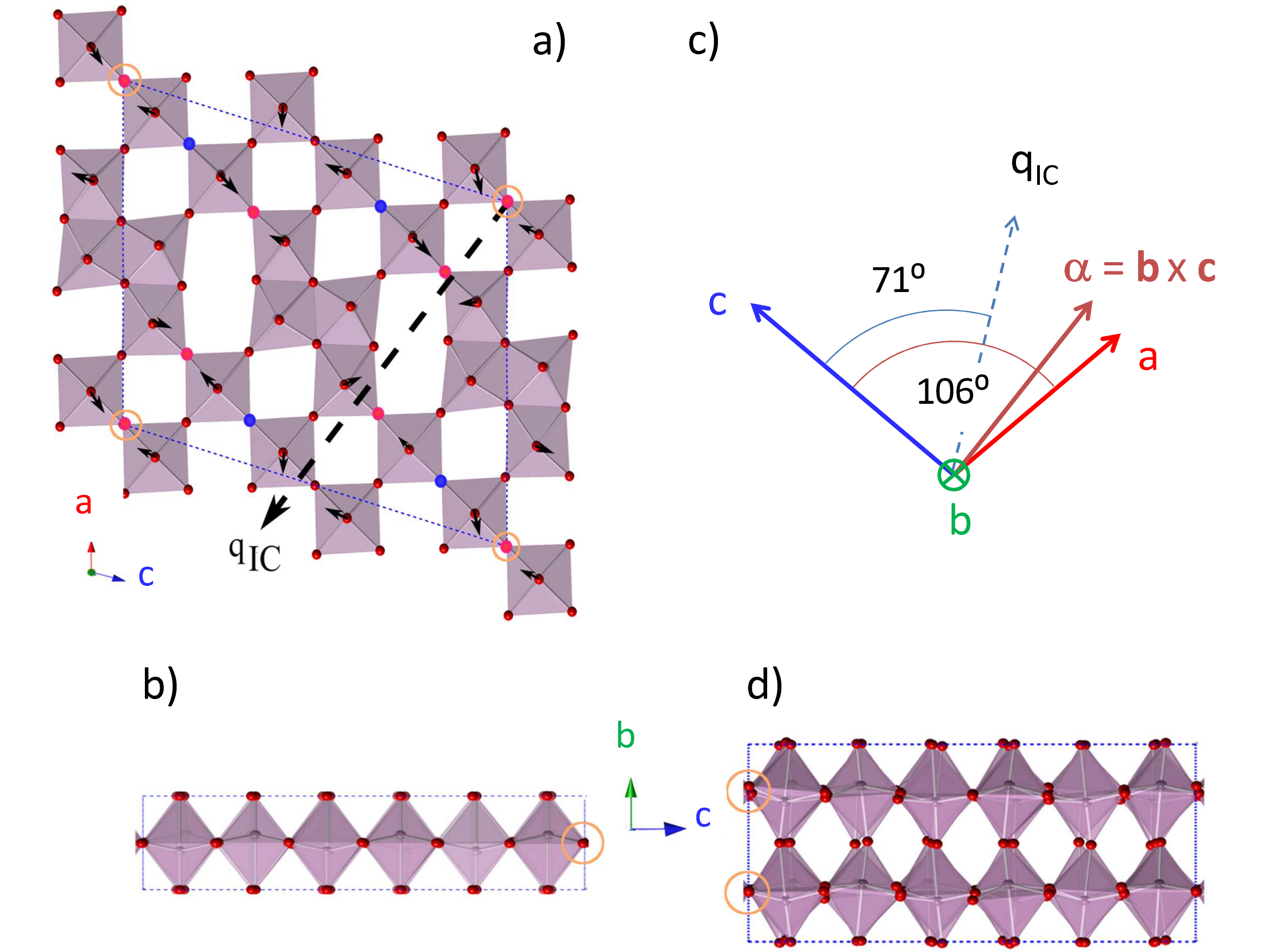} \caption{The crystal structure of Mo$_{8}$O$_{23}$ after ref.~\cite{Sato,Fujishita87}.
The fine-dashed lines denote the unit cells and the encircled atoms
the inversion centers of the high-$T$ structure. a) The high-$T$
structure in the $ac$-plane ($T>T_{\mathrm{ICDW}}\sim350$
K). Arrows indicate the directions of the low-$T$ CDW displacements
for the top layer of the low-$T$ bilayer structure. The blue (red)
dots indicate negative (positive) out-of-plane displacements. b) The
high-$T$ structure in the $bc$-plane. c) The coordinate
system notation used in optical measurements. The dashed arrow ($q_{\mathrm{IC}}$)
indicates the $ac$-plane projection of the CDW wave vector \cite{Sato}.
d) The low-$T$ bilayer structure in the $bc$-plane. }
\label{fig:structure} 
\end{figure}

In the case of Mo$_{8}$O$_{23}$, structural \cite{Sato,Fujishita87,Fujishita,Komdeur}
and transport studies \cite{Sato} suggest that following the emergence
of a high-temperature incommensurate order at
$T_{\mathrm{IC}}\sim350$\,K, an intriguing and unusual C$_{2}$-and
inversion symmetry breaking commensurate charge density wave (CDW)
state (see Fig.~\ref{fig:structure}) occurs below $T_{\mathrm{IC-C}}=285$\,K.
Curiously, the resistivity shows an anomalous peak around $T_{\mathrm{el}}\sim150$\,K
that cannot be understood in terms of the high-temperature CDW ordering.
Thus the nature of the low-temperature ($T<T_{\mathrm{IC-C}}$) phase
seems to be more complicated than anticipated, possibly with multiple
orders present.

CDW's that break not only translational but also rotational in-plane
symmetries of the underlying lattice are rare \cite{TbTe3KapitulnikSTM}.
On the other hand, the antiferrodistortive structural transitions
driven by soft zone boundary phonon with displacements described by
octahedral rotation are common phenomena in the insulators and could
produce incommensurate phases \cite{axe1986phase} due to the competition
of short-range forces \cite{Bak1982review}. Such a scenario has been
proposed also for \mox \cite{Schlenker1}, supported by the first
band structure consideration \cite{Whangbo} that found only a weak
band dispersion along the commensurate wave vector $q_{\mathrm{C}}=(0,0.5,0)$.
The increase of $T_{\mathrm{IC-C}}$ under pressure \cite{Sowa} is
also common to the antiferrodistortive transitions \cite{Samara,SAMARA19821}
and not expected for a CDW-type nesting instability \cite{Monceau},
although some exceptions were reported \cite{PressureZrTe3}.

Yet room-temperature STM atomic-resolution images of \mox ~{[}cleaved
(010) surface{]} demonstrate charge modulation \cite{STMMo8O23},
characteristic for CDW's \cite{Monceau}. Moreover, the density-functional-theory
(DFT) ab-initio calculations (to be reported elsewhere \cite{DOSpaper})
show enhancement of the Lindhard susceptibility in a broad range of
wave vectors, much like in TiSe$_{2}$ \cite{CDWTiSe2}, and give
some support for the nesting-driven CDW scenario for both, the commensurate
and incommensurate phases.

In the present case the equilibrium-state experimental techniques
may be insufficient to completely disentangle structural and electronic
orders and the origin of the underlying interactions and need to be
complemented by non-equilibrium techniques. Among the latter the ultrafast
time-resolved optical spectroscopy is useful in elucidating the single
particle and collective excitations in relation to the various order
parameters by separating their timescales and may even help to detect
electronic gaps \cite{DemsarBlueBronze,DemsarHeavyFermions2006,YusupovNature,TomazPRLpnictides,DemsarPRB,OrganicToda2011,Hsieh2018}.
Moreover, the probe-polarization resolved pump-probe studies can help
to identify different components of the transient response in non-centrosymmetric
materials \cite{GaAsoffdiagonal} or different orders in correlated
systems such as pseudogaps and superconducting gaps in the cuprates
\cite{Toda1} and organic superconductors \cite{TodaTsuchiya,Nakagawa2016,Tsuchiya2017}.

Here we present a probe-polarization-resolved transient reflectivity
study of the phase diagram and particularly the low-temperature anomalous
CDW state in single crystals of Mo$_{8}$O$_{23}$, aiming to disentangle
various order parameters contributions to the relaxation and get insight
into their respective symmetries.

We found a very anisotropic transient response that develops gradually
below $T_{\mathrm{IC}}$ and saturates below $T_{\mathrm{el}}\sim150$\,K.
We observe a prominent coherent mode (CM) arising around $T_{\mathrm{IC}}$
with an unusual non-monotonic temperature dependence of frequency
and damping. A few less intense coherent modes become visible below
$T_{\mathrm{el}}$. In addition, a sub-ps exponential relaxation mode
(EM) is present at all temperatures. While the appearance of EM and
CM modes is a typical CDW transient reflectivity phenomenon, in the
present case the decay-time of the exponential EM and the damping
and frequency of the coherent CM demonstrate extrema around $T_{\mathrm{el}}$,
far from the IC and C transitions. Overall, we find that the transient
response is consistent with the presence of a (pseudo)gap in the electronic
spectrum below $T_{\mathrm{IC}}\sim350$\,K with an additional gap
opening below $\sim160\,\mathrm{K}\sim T_{\mathrm{el}}$. At the same
time, the behavior of the CM is quite unconventional suggesting that
the EM and CM are not coupled to the IC(-C) CDW order parameter only.

\section{Experimental and theoretical methods}

\begin{figure*}[t]
\includegraphics[width=1\columnwidth]{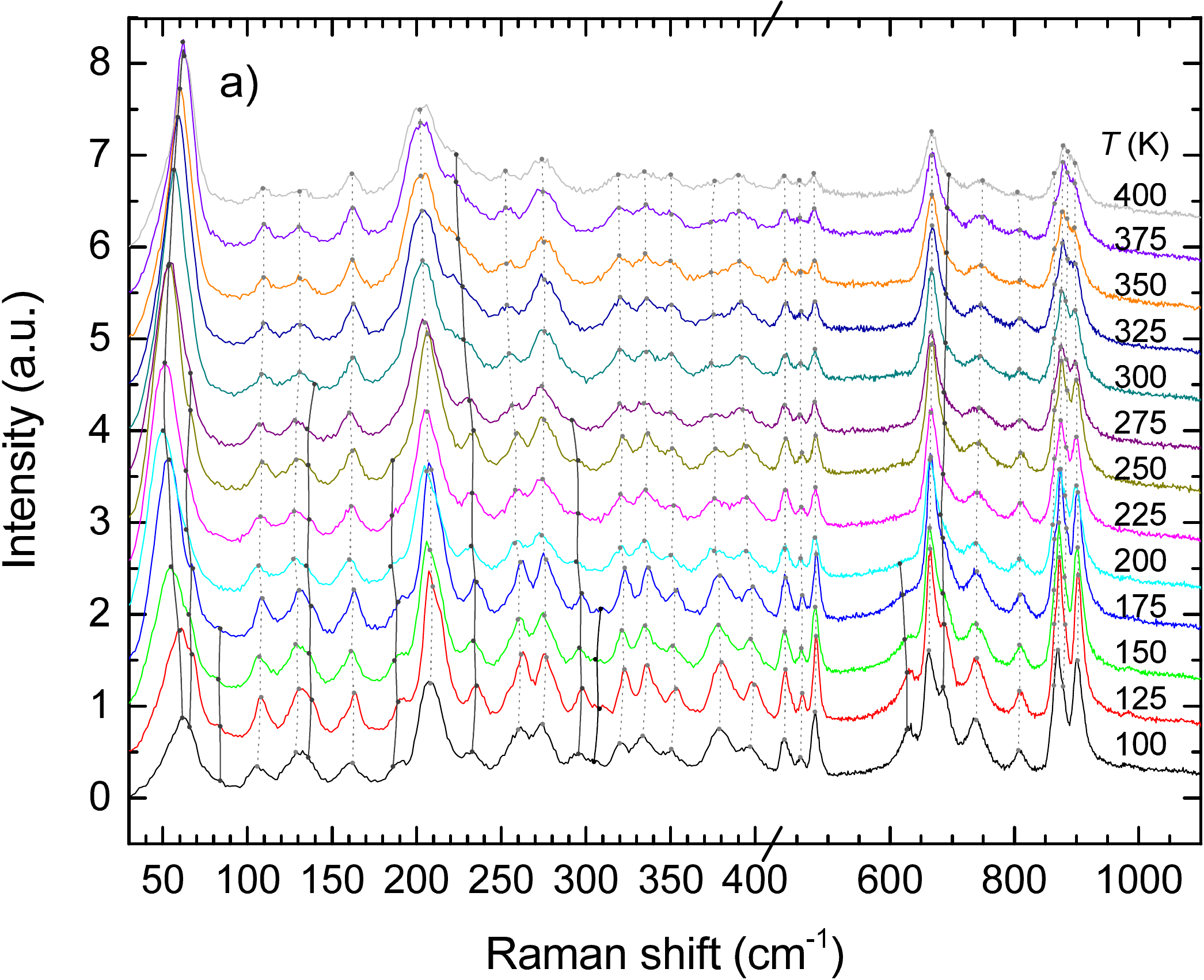} \hspace{0.05\columnwidth}\includegraphics[clip,width=0.9\columnwidth]{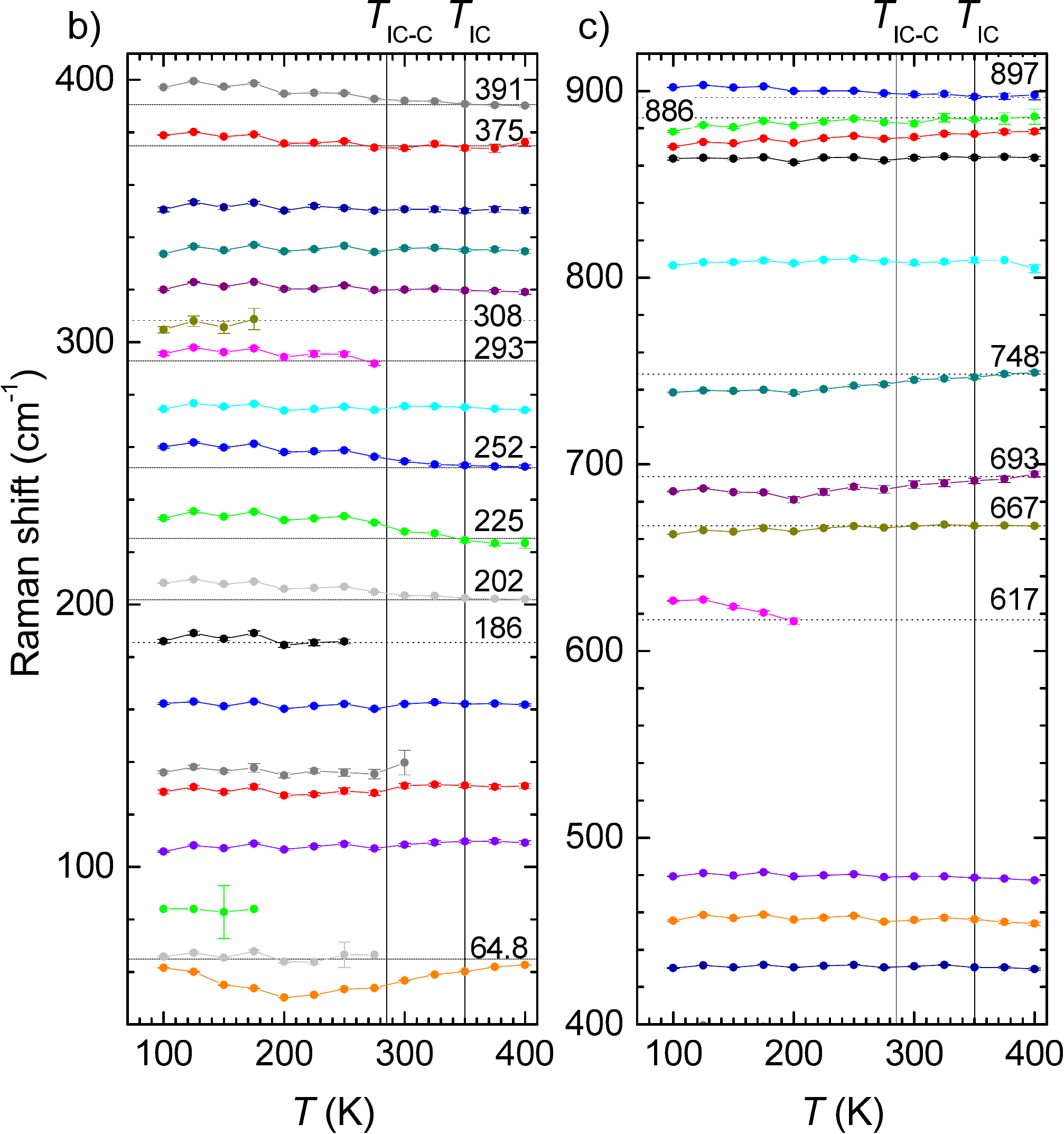}
\caption{(a) Temperature
dependent vibrational Raman spectra. The full vertical lines indicate
the modes that show more significant $T$-dependent intensity. (b)
and (c) Temperature
dependence of the peak positions obtained by means of multiple Lorentzian-peak
fits.}
\label{fig:Raman} 
\end{figure*}

The samples were grown as described in the literature \cite{Mo8O23growth,DOSpaper}.
The single-crystal nature of the samples was confirmed by XRD. The
resistivity of the samples is consistent with previous reports \cite{Sato}
and exhibits a characteristic peak at $T_{\mathrm{el}}\sim150$\,K
\cite{DOSpaper}. The orientation of the crystal axes was determined
by means of the X-ray diffraction (XRD) and the polarized optical
microscopy.

The transient reflectivity studies were performed using a standard
pump-probe setup using 50\,fs linearly polarized laser pulses at
800\,nm wavelength and the 250\,kHz repetition rate as presented
in detail elsewhere \cite{Mertelj2017}. The probe-photon energy was
at the laser fundamental, $\hbar\omega=1.55$\,eV, while the pump-photon
energies were either 1.55\,eV or 3.1\,eV. The transient reflectivity,
$\Delta R/R$, was measured at a near-normal incidence from the $ac$-plane
with the incident and reflected light propagating at $\sim15$ degrees
with respect to the crystal $b$-axis. The angle was slightly different
in different runs with no noticeable effect on the results. No dependence
of the signal on the pump polarization was found with the 3.1\,eV
pump-photon energy.
The beam diameters at the sample surface were $\sim100$\,$\mu$m
and $\sim50$\,$\mu$m for the pump and the probe beams, respectively.
The low fluences of the pump $F_{\mathrm{pu}}<=40$\,uJ/cm$^{2}$
were verified to transiently increase lattice temperature for less
than 5 K. This agrees with an estimate based on the penetration depth
of 20\,nm from ab initio \cite{DOSpaper} calculations.
Samples were cleaved before experiments, but no major difference in
the response from the cleaved and as-grown $ac$ crystal surfaces
was observed. The temperature dependence of $\Delta R/R$ was measured
with the pump fluence of 10\,uJ/cm$^{2}$ at a fixed linear probe
polarization oriented approximately along the maximum of the low-$T$
anisotropic $\Delta R$.
The polarization dependence of $\Delta R/R$ at selected temperatures
was measured using the fluence of $\sim40$\,uJ/cm$^{2}$ with the
probe fluence being $\sim8$ times lower. In this range, the $\Delta R/R$
transients were found to scale linearly with the fluence. The characteristic
changes of $\Delta R/R$ with the probe-polarization when varying
temperature over the IC-C transition were always measured in a single
run with a constant probe incidence angle.
To measure the static
reflectivity anisotropy, $R(\theta)$, the angular dependence of either
the reflected 1.55-eV probe-beam intensity or the reflected broadband
supercontinuum probe-beam spectral density was measured. In the later
case no in-situ reference mirror was used so the magnitude-calibrated
$R(\theta)$ is available only at 1.55-eV photon energy.

Temperature dependent Raman spectra were measured in the backscattering
geometry using 10x microscope objective. The excitation laser wavelength
was 632.8\,nm with the incident laser power of a few hundred\,$\mu$W.

\section{Results}

\subsection{Raman}

In Fig. \ref{fig:Raman} we show temperature dependent Raman spectra
measured in the backscattering geometry along the $b$ axis, with
the analyzer parallel to the laser polarization. The orientation of
the crystal in the $ac$ plane was chosen for the largest intensity
in the low frequency region.

A large number of vibrational modes is observed. No clear new modes
appear at $T_{\mathrm{IC}}$. Below $T_{\mathrm{IC-C}}$ a few modes
(at 137, 187, 233 and 293\,cm$^{-1}$) develop to well defined peaks
from broad shoulders above $T_{\mathrm{IC-C}}$. Another mode
at 617\,cm$^{-1}$ with a significant $T$-dependent frequency shift
rises below $175-200$\,K. The other modes
show slight increase of intensity with decreasing $T$. Below $175-200$\,K
some of these modes (the 478\,cm$^{-1}$ mode for example) show an
enhanced intensity increase.

The most striking feature of the spectra is the temperature dependence
of the lowest frequency $\sim65$\,cm$^{-1}$ mode that shows a strong
softening with decreasing $T$ down to $175-200$\,K
followed by hardening back to the high-$T$ frequency upon further
cooling. Among the other modes some show hardening, some softening,
and some also no appreciable $T$ dependence.

Most of the shifting modes show the largest frequency shifts across
$T_{\mathrm{IC-C}}$ with an onset of shift observable already below
$T_{\mathrm{IC}}$. A marked exception is the mode at 375\,cm$^{-1}$
that shows a frequency shift below $175-200$\,K only.
Note that the peaks that demonstrate changes below 200\,K
are rather weak and the changes are clearly observed only below $\sim175$\,K. Thus
200\,K is a guideline temperature corresponding to the maximal observed softening
of the anomalous mode at the $\sim65$\,cm$^{-1}$.

\subsection{Reflectivity anisotropy}

\begin{figure}[ht]
\includegraphics[width=0.8\columnwidth]{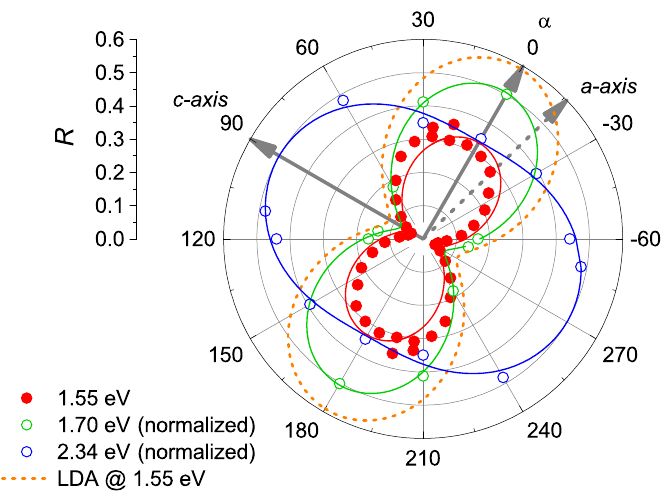} \caption{Anisotropy of the static reflectivity, $R$, at different photon energies.
The data represented with open symbols are normalized to $R_{\mathrm{max}}(\theta)=0.5$ due to the lack
of calibration. The continuous lines represent the fits using (Eq.~\ref{eq:Rangular}).}
\label{fig:Raniso} 
\end{figure}

\begin{figure*}
\includegraphics[width=1\textwidth]{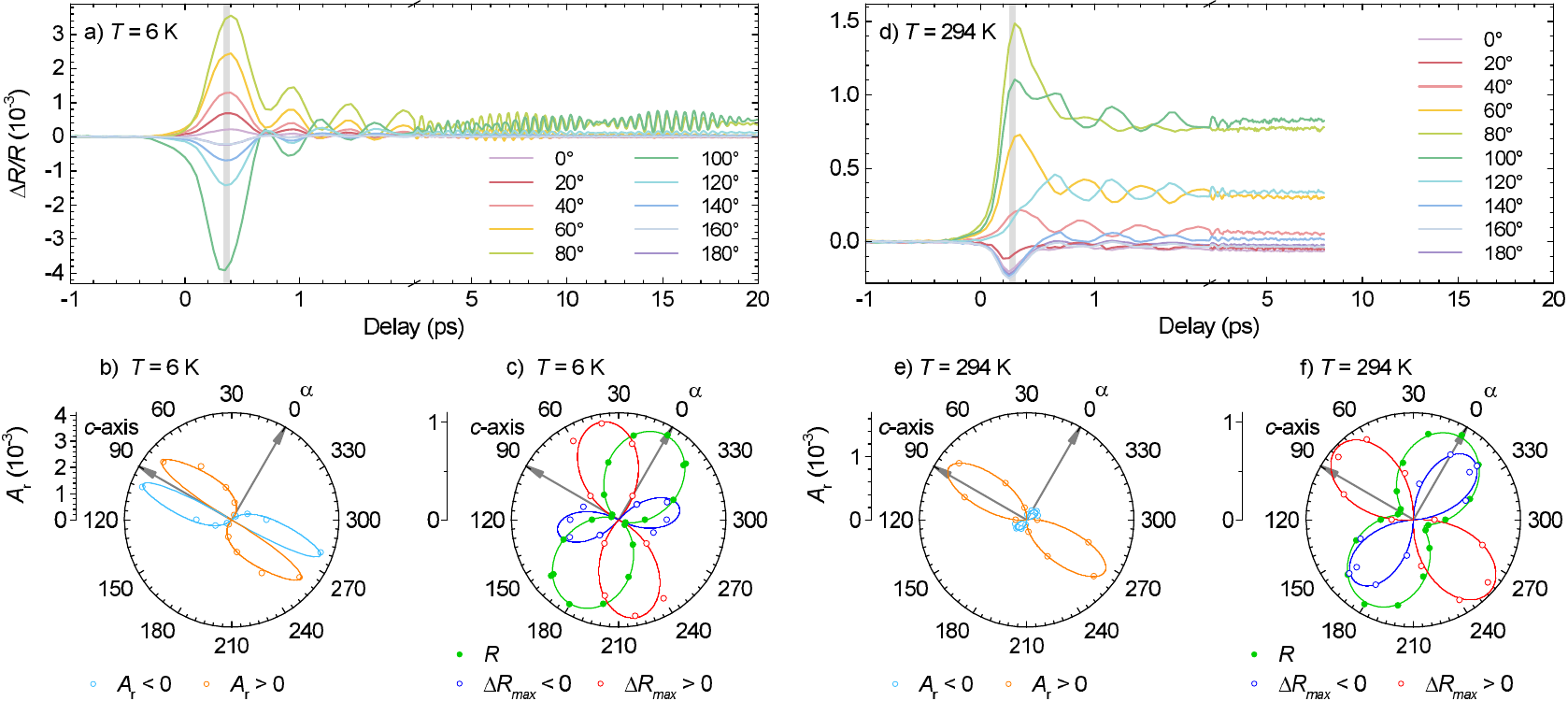}\caption{(a), (d)
The transient reflectivity, $\Delta R/R$, as a function of the probe
polarization angle at 6 K and 294 K respectively. (b),
(e) Angular dependence
of the transient reflectivity amplitude, $A_{\mathrm{r}}$,
at 6 K and 294 K respectively. (c),
(f) Comparison of the
unnormalized transient reflectivity amplitude, $\Delta R_{\mathrm{max}}$,
to the static reflectivity, $R$ (the
green curve with the
maximum at zero angle) at 6 K and 294 K respectively.
The blue colors in (b), (c),
(e)
and (f) represent
the negative lobes. \label{fig:figDRAniso}}
\end{figure*}

A low monoclinic symmetry implies that one can expect a strongly anisotropic
optical response so in order to analyze the transient-reflectivity
polarization dependence we first need to determine the anisotropy
of the static optical reflectivity at the relevant wavelengths used
in the experiments. In Fig.~\ref{fig:Raniso} we show the \emph{static} $ac$-plane reflectivity $R$ as a function of the polarization angle.
A strong angular dependence of $R$ at 1.55\,eV and 1.70\,eV is
found that becomes weaker at the highest photon energy 2.34\,eV.
To analyze this in more detail, we recall that the monoclinic symmetry
of \mox crystals \cite{Sato,Komdeur} implies \cite{Yariv} that
the dielectric tensor has the form 
\begin{equation}
\hat{\epsilon}=\left(\begin{array}{ccc}
\epsilon_{\mathrm{\alpha\alpha}} & 0 & \epsilon_{\mathrm{\alpha c}}\\
0 & \epsilon_{\mathrm{bb}} & 0\\
\epsilon_{\mathrm{\alpha c}} & 0 & \epsilon_{\mathrm{cc}}
\end{array}\right),\label{eq:eps}
\end{equation}
where $\epsilon_{ij}$ represent the complex dielectric susceptibility
components in the orthogonal coordinate system, $\{\vec{\alpha}=\vec{b}\times\vec{c},\vec{b},\vec{c}\}$,
where $\vec{a}$,$\vec{b}$ and $\vec{c}$ correspond to the crystal-axes
directions (see Fig. \ref{fig:structure}). This can in general yield
complex eigenvectors of the dielectric tensor in the $ac$-plane that
are wavelength-dependent. As a result the general $ac$-plane complex
reflectivity is given by \cite{Koch1974}

\begin{equation}
\hat{r}_{\mathrm{ac}}=\left(\begin{array}{cc}
r_{\mathrm{\alpha\alpha}} & r_{\mathrm{\alpha c}}\\
r_{\mathrm{\alpha c}} & r_{cc}
\end{array}\right).\label{eq:r-tensor}
\end{equation}
Under a polarized microscope the \mox single crystals show good cross-polarized
extinction with the incident polarization along the $\alpha$ and
$c$ axes (within the experimental error of $\sim\pm5^{\circ}$) implying
that $|r_{\mathrm{\alpha c}}|\ll|r_{cc}|<|r_{\alpha\alpha}|$. The
observed dependence can therefore be fit with 
\begin{equation}
R(\theta)=C_{1}+C_{2}\cos(2\theta),\label{eq:Rangular}
\end{equation}
where $\theta$ is the angle relative to the $\alpha$ axis.

In Fig.~\ref{fig:Raniso} we also show for comparison the reflectivity
anisotropy calculated by means of the DFT \cite{DOSpaper} for Mo$_{8}$O$_{23}$
in the high-temperature structure. Overall, the calculations yield
a similar anisotropy as the observed one, but the magnitude of the
calculated reflectivity is approximately a factor of 2 larger than
the measured values. Overall we believe that discrepancies
are within the accuracy of DFT (LDA) calculations. For the
photon energies other than 1.55\,eV the DFT (LDA) calculations also demonstrate qualitatively the same anisotropies
as experimental ones.

\begin{figure*}[t]
\includegraphics[width=0.85\columnwidth]{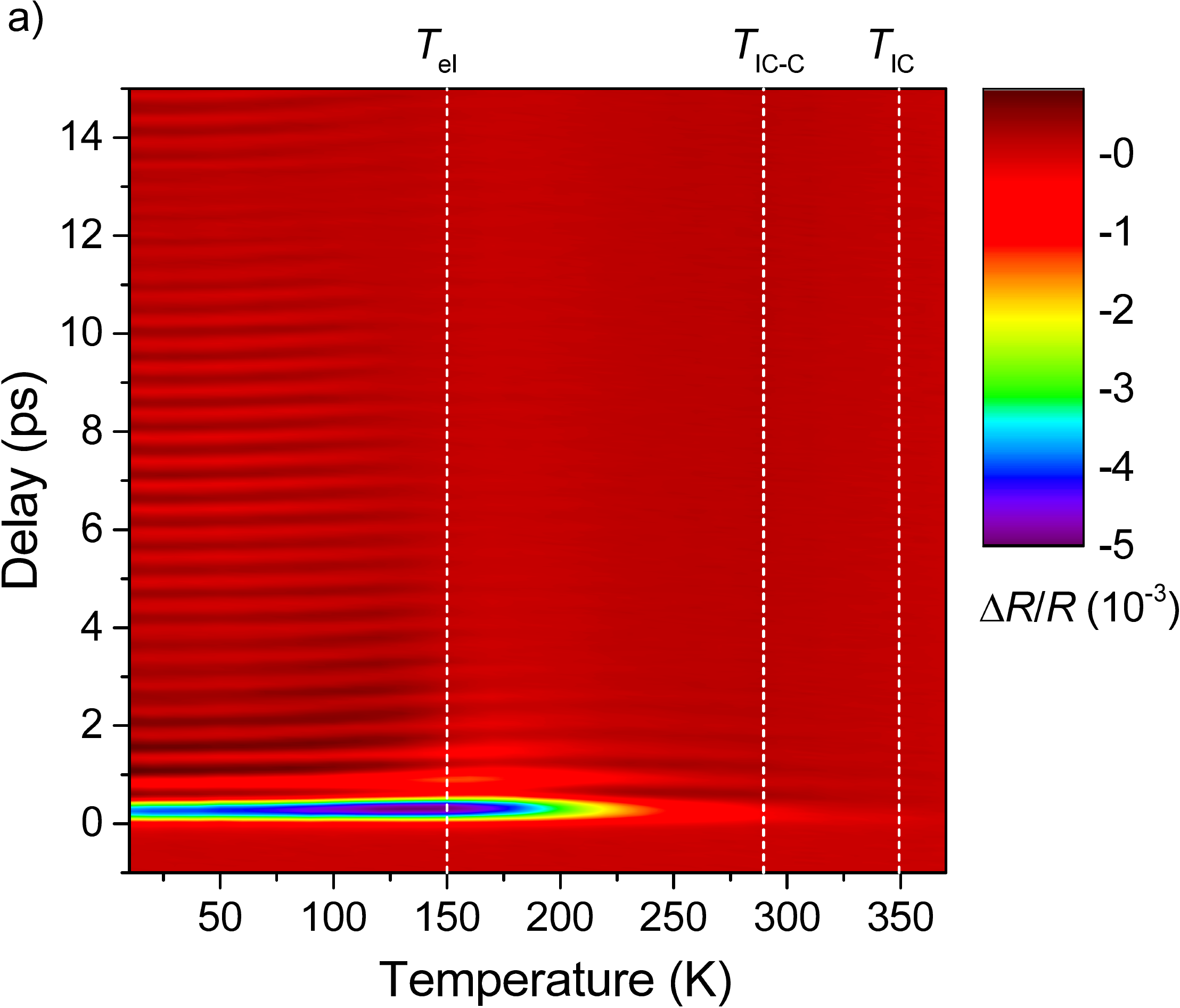}\hspace{0.15\columnwidth}\includegraphics[width=0.85\columnwidth]{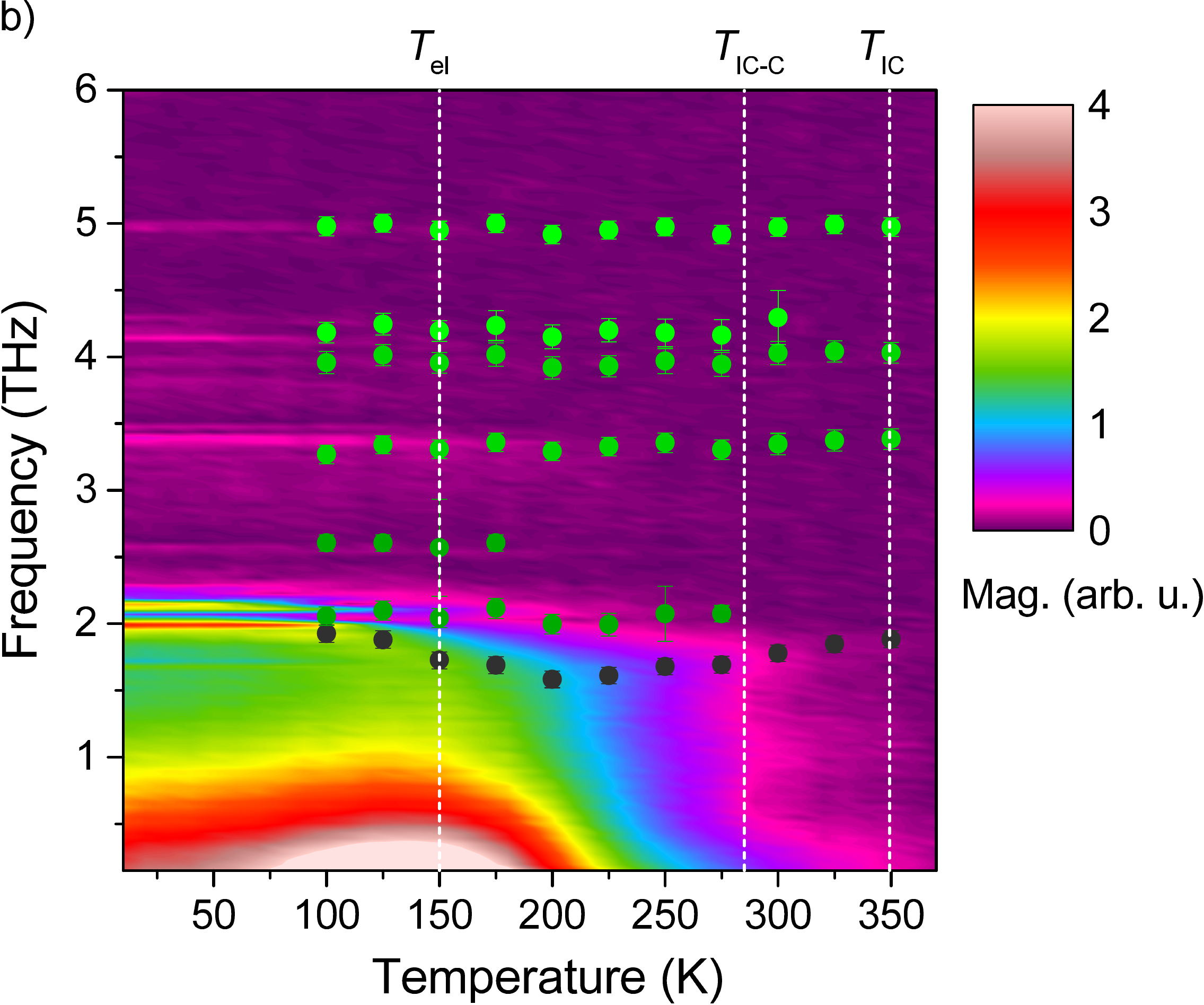}
\caption{(a) Temperature dependence of the transient reflectivity measured
at the 10\,$\mu$J/cm$^{2}$ pump fluence and $\theta\sim40{}^{\circ}$.
(b) The Fourier transform of the data presented in (a). Full symbols
correspond to the frequencies of the low frequency Raman peaks. The
vertical dashed lines indicate the transition temperatures: $T_{\mathrm{IC}}\sim350$\,K
and $T_{\mathrm{IC-C}}=285$\,K and $T_{\mathrm{el}}$.}
\label{fig:tdep} 
\end{figure*}

\begin{figure*}[t]
\includegraphics[width=0.9\textwidth]{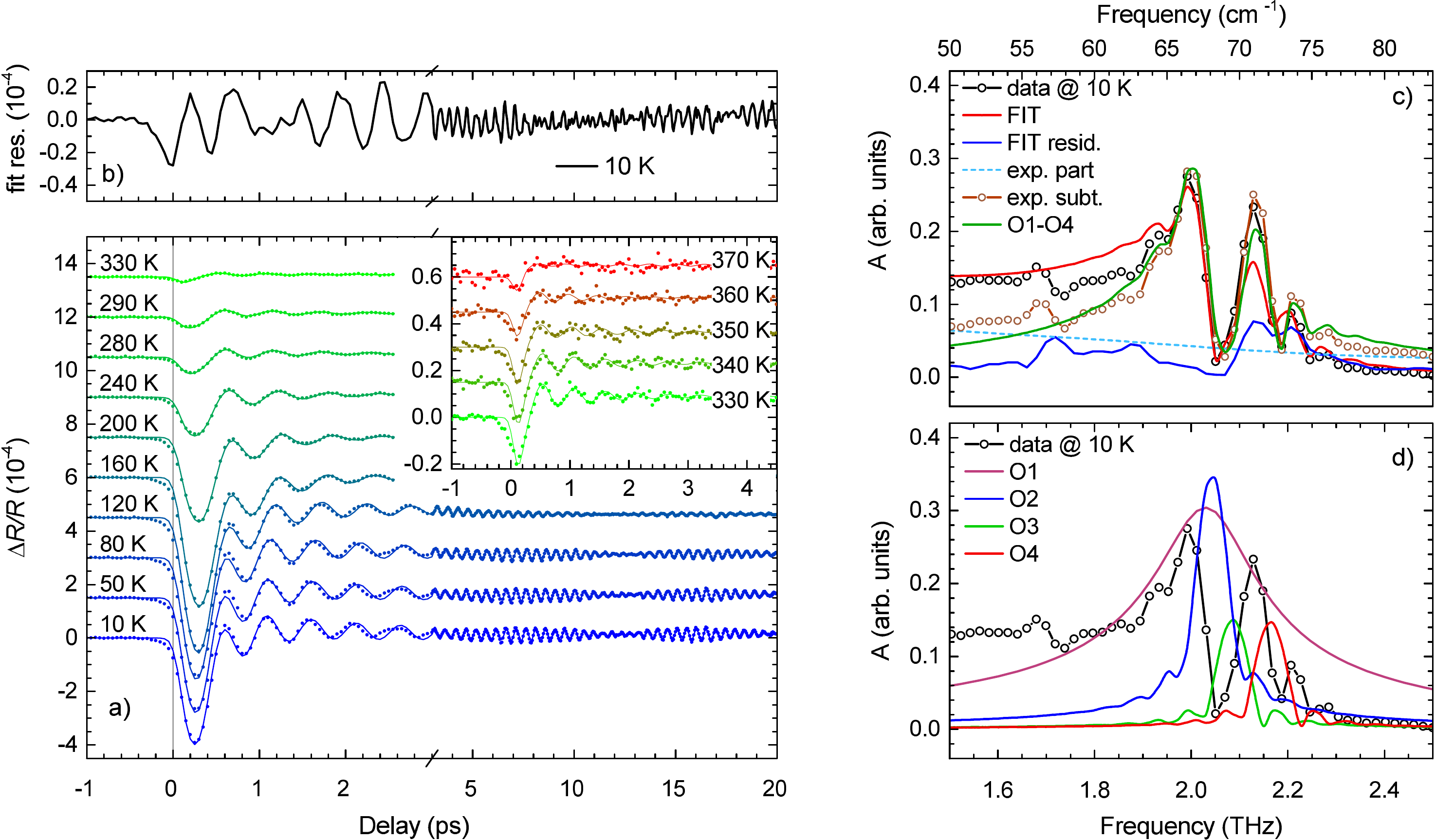} \caption{(a) DCE model (Eq.~\ref{eq:fitfunc}) fits (lines) to the transient-reflectivity
data (circles). The high-$T$ data are shown in the inset. The full decomposition of the raw data at $T=10$\,K into the components
is shown in Appendix. (b) The fit residuum at $T=10$\,K reveals
the weaker high frequency modes. (c) The Fourier transform amplitude
of the transients and fit components at $T=10$\,K. (d) The $T=10$\,K
Fourier transform amplitude of the four fit coherent modes compared
to the data. The lineshapes of the two narrowest coherent modes are
due to the finite scan-length. The side lobes of all the modes are
also a consequence of the finite scan-length.}
\label{fig:4ores} 
\end{figure*}

\subsection{Transient reflectivity: anisotropy}

The low $T$ transient reflectivity, $\Delta R/R$ shown in Fig. \ref{fig:figDRAniso}
(a) is dominated by the coherent oscillatory response. A clear beating
of the oscillatory responses indicates the presence of at least two
modes.

The angular dependence of the $\Delta R/R$ amplitude, $A_{\mathrm{r}}$,
shown in Fig \ref{fig:figDRAniso} (b) and (e) shows a peculiar angular
dependence. The
unnormalized $\Delta R_{\mathrm{max}}$ shows a phase-shifted angular
dependence of the similar form (Eq.~\ref{eq:Rangular}),
$\Delta R_{\mathrm{max}}\sim1+C_{3}\cos(2[\theta-\theta_{0}])$, shown
in Fig. \ref{fig:figDRAniso} (c) and (e). The peculiar shape is therefore
a result of the angular phase shift, $\theta_{0}\sim40{}^{\circ}$
at $T=6$ K, of $\Delta R_{\mathrm{max}}(\theta)$ with respect to
$R(\theta),$

\begin{equation}
A_{\mathrm{r}}\sim\frac{\Delta R_{\mathrm{max}}(\theta)}{R(\theta)}\sim\frac{1+C_{3}\cos(2[\theta-\theta_{0}])}{1+\frac{C_{2}}{C_{1}}\cos(2\theta)}.
\end{equation}

The $A_{\mathrm{r}}$ angular dependence shows a handedness that was different in different crystals, but was never
observed to change upon temperature cycling. This is consistent with the symmetries of both the high-$T$
($P2/c$) and the low-$T$
($Pc$) crystal space group that allow a pseudovector along the $b$-axis.
Different handedness in different crystals can be related to different
orientation of the particular crystal pseudovector relative to the
light propagation direction.

\subsection{Transient reflectivity: temperature dependence}

The temperature dependence of the transient reflectivity was measured
at $\theta\sim40^{\circ}$ where the low-$T$ $\Delta R_{\mathrm{max}}$
shows the maximum. The transient reflectivity is shown in Fig.~\ref{fig:tdep}(a).
One can see clear coherent oscillations that appear soon after cooling
below $T_{\mathrm{IC}}$. Also on cooling, the amplitude of the transient
reflectivity on short timescales ($<1$\,ps) begins to increase.
Similar results were reported for other CDW compounds \cite{DemsarBlueBronze,YusupovPRL101},
including Mo$_{4}$O$_{11}$,
another member of the
MoO$_{3-x}$ family \cite{Milos}.

The corresponding Fourier transformation of the data is shown in {[}Fig.
\ref{fig:tdep} (b){]}. One can see several modes with the most intensive
group centered around 2\,THz. These modes are strongly damped above
250\,K. The higher frequency modes are rather weak with narrow linewidths
and are clearly observed only at low $T$.

To analyze the signal we fit the data using the displacive coherent
excitation model \cite{zeiger1992theory} (DCE) where the transient
reflectivity is given by 
\begin{gather}
\frac{\Delta R}{R}=(A\mathrm{_{displ}}-\sum A_{\mathrm{O}i})\int_{0}^{\infty}G(t-u)e^{-u/\tau_{\mathrm{displ}}}du\nonumber \\
+\sum A_{\mathrm{O}i}\int_{0}^{\infty}G(t-u)e^{-\gamma_{i}u}[\cos(\Omega_{i}u)\nonumber \\
-\beta_{i}\sin(\Omega_{i}u)]du\nonumber \\
+\sum_{j\in\{1,2\}}A_{\mathrm{e}j}\int_{0}^{\infty}G(t-u)e^{-u/\tau_{j}}du,\label{eq:fitfunc}
\end{gather}
where $\beta_{i}=(1/\tau_{\mathrm{displ}}-\gamma_{i})/\Omega_{i}$
and $G(t)=\sqrt{\nicefrac{2}{\pi}}\tau_{\mathrm{p}}\exp(-2t^{2}/\tau_{\mathrm{p}}^{2})$
with $\tau_{\mathrm{p}}$ being the effective pump-probe pulse cross-correlation
width. In the DCE model the coherent modes are driven with an exponentially
relaxing displacive mode (ERDM) with the relaxation time $\tau_{\mathrm{displ}}$
and the amplitude, $A\mathrm{_{displ}}$. $A_{\mathrm{O}i}$, $\Omega_{i}$,
$\gamma_{i}$ are the oscillating modes amplitudes, frequencies and
damping factors, respectively, while $A_{\mathrm{e}j}$ and $\tau_{j}$
are the amplitudes and relaxation times of two additional exponentially
relaxing modes that are necessary to fit the data: component
e1 that describes the
picosecond-scale
relaxation and component
e2 which accounts for the
long-lived ($>1$\,ns) offset (see Appendix for the decomposition
of the raw data at 10\,K into the components). To reduce the number
of parameters we ignore the weak and featureless high frequency modes
and focus on the strongest group of modes around $\sim2$\,THz.

In Fig.~\ref{fig:4ores}(a) we show the resulting fits along the
measured data. Overall, the fit is good, as seen also from the Fourier
transform shown in Fig.~\ref{fig:4ores}(c), and small fit residuals,
Fig.~\ref{fig:4ores}(b). For this high quality of the fit, in particular
to reproduce the double-peak structure around $\sim2$\,THz, four
oscillatory modes labeled $\mathrm{O1-O4}$ in Table~\ref{tab:modes}
are necessary.

\begin{figure}[ht!]
\includegraphics[width=0.75\columnwidth]{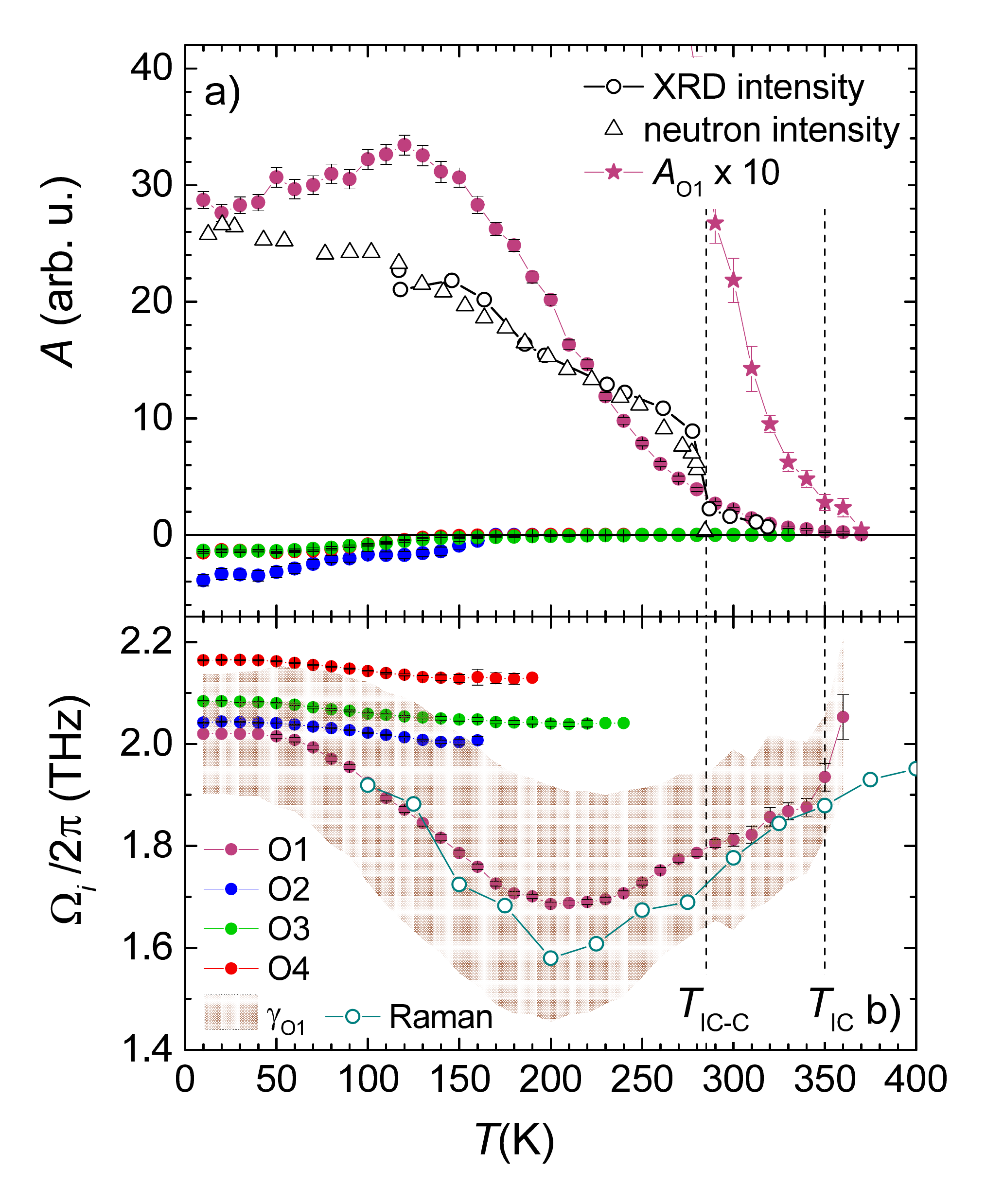} \caption{(a) Temperature dependence of the coherent mode
amplitudes compared
to the structural order parameters from \cite{Sato}. Temperature dependence of the coherent mode
frequencies
(b). The open symbols correspond to the Raman scattering 65-cm$^{-1}$
mode frequency (mode RO1). The line-width of the dominant coherent
mode $\mathrm{O1}$ is shown by the shaded area.}
\label{fig:CM} 
\end{figure}

Let us describe the temperature dependence of those oscillatory modes.
Fig.~\ref{fig:CM} shows the temperature dependence of the fitted
frequencies and the amplitudes. On cooling the most prominent mode
$\mathrm{O1}$ appears just above $T_{\mathrm{IC}}\sim350$\,K with
a strong increase of the amplitude below $\sim300$ K. The mode frequency
and rather large damping (linewidth) strongly depend on temperature.
Interesting non-monotonic behavior is observed: upon decreasing $T$
the mode first softens from $\Omega_{1}/2\pi\sim2.1$\,THz to the
minimum frequency of $\sim1.7$\,THz at $T_{\mathrm{m}}\sim200$\,K
and then hardens with further cooling to 2.02\,THz. The mode damping
shows a broad maximum ($\gamma_{\mathrm{max}}/2\pi\sim0.24$\,THz)
spread between $\sim200$\,K and $\sim100$\,K.

\begin{table}
\begin{tabular}{c|c|c|c}
 & $\omega/2\pi$ (THz)  & $\gamma/2\pi$ (THz)  & $A$ (arb. u.)\tabularnewline
\hline 
O1  & 2.02  & $0.1$2  & $28.4$\tabularnewline
O2  & 2.04  & 0.007  & $-5.7$\tabularnewline
O3  & 2.09  & $<0.007$  & $-1.5$\tabularnewline
O4  & 2.17  & $<0.007$  & $-1.4$\tabularnewline
\end{tabular}\caption{The four strongest coherent mode fit parameters at $T=10$ K.\label{tab:modes}}
\end{table}

\begin{figure}[ht!]
\includegraphics[width=0.75\columnwidth]{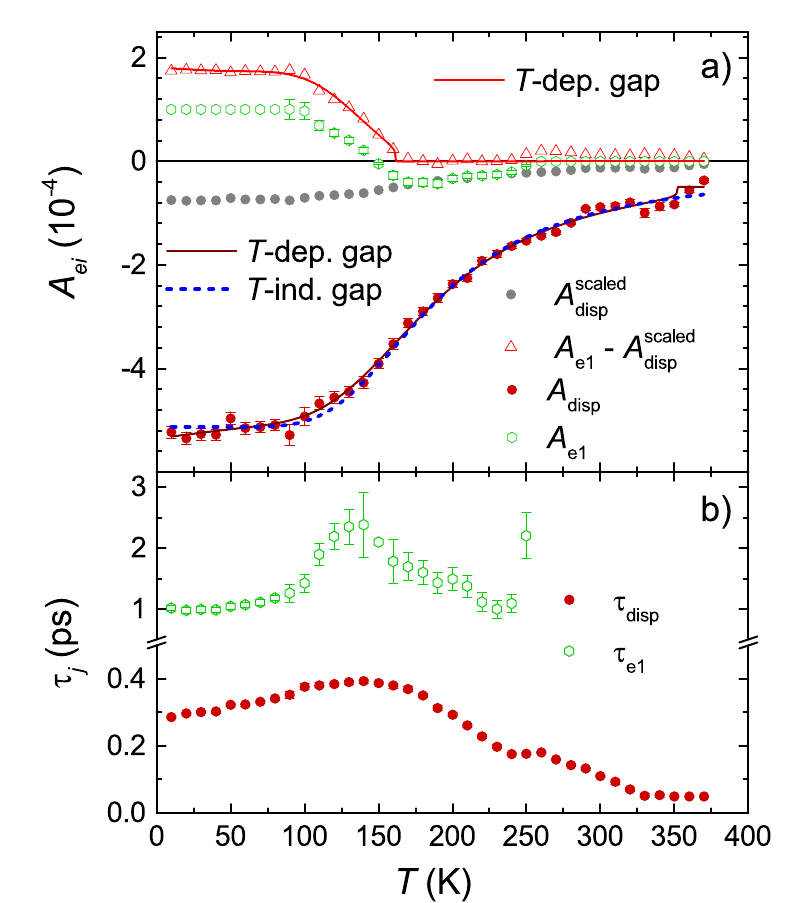} \caption{Temperature dependence of the two fast non-oscillating component parameters:
(a) the amplitudes and (b) the corresponding decay times. The lines
in (a) show the bottleneck-model fits with the temperature-independent
gap (Eq.~\ref{eq:Tindepbottleneck}) (dashed line) and the temperature-dependent
BCS gap (Eq.~\ref{eq:Tdepbottleneck}) (full lines). See text for
more details.}
\label{fig:AvsT} 
\end{figure}

The three additional fitted modes $\mathrm{O2-O4}$ become observable
in the 250-170\,K range, depending on the mode, and show smaller
damping (see Table \ref{tab:modes}) and weaker $T$ dependence (Fig.~\ref{fig:CM}).
The minor higher frequency modes (above 2.5\,THz) also appear at
different temperatures below $170-200$\,K with a general
trend of the higher frequency modes appearing at lower $T$.

In Fig.~\ref{fig:AvsT} we show the amplitudes of two main non-oscillatory
components, the ERDM ($A_{\mathrm{displ}}$) and the picosecond
component e1 ($A_{e1}$) as well as the corresponding relaxation
times. The ERDM is always present and its amplitude dominates the
signal in the full temperature range. Its amplitude and the relaxation
time show a smooth weak evolution across $T_{\mathrm{IC}}$ and the
lock-in transition at $T_{\mathrm{IC-C}}$. Below $\sim250$\,K weaker
picosecond component e1 appears. Its amplitude changes sign close
to 150\,K. Close to this temperature both relaxation times are maximal.

\subsection{Transient reflectivity: T-dependent anisotropy and evidence of twinning.}

\begin{figure}[ht!]
\includegraphics[width=0.9\columnwidth]{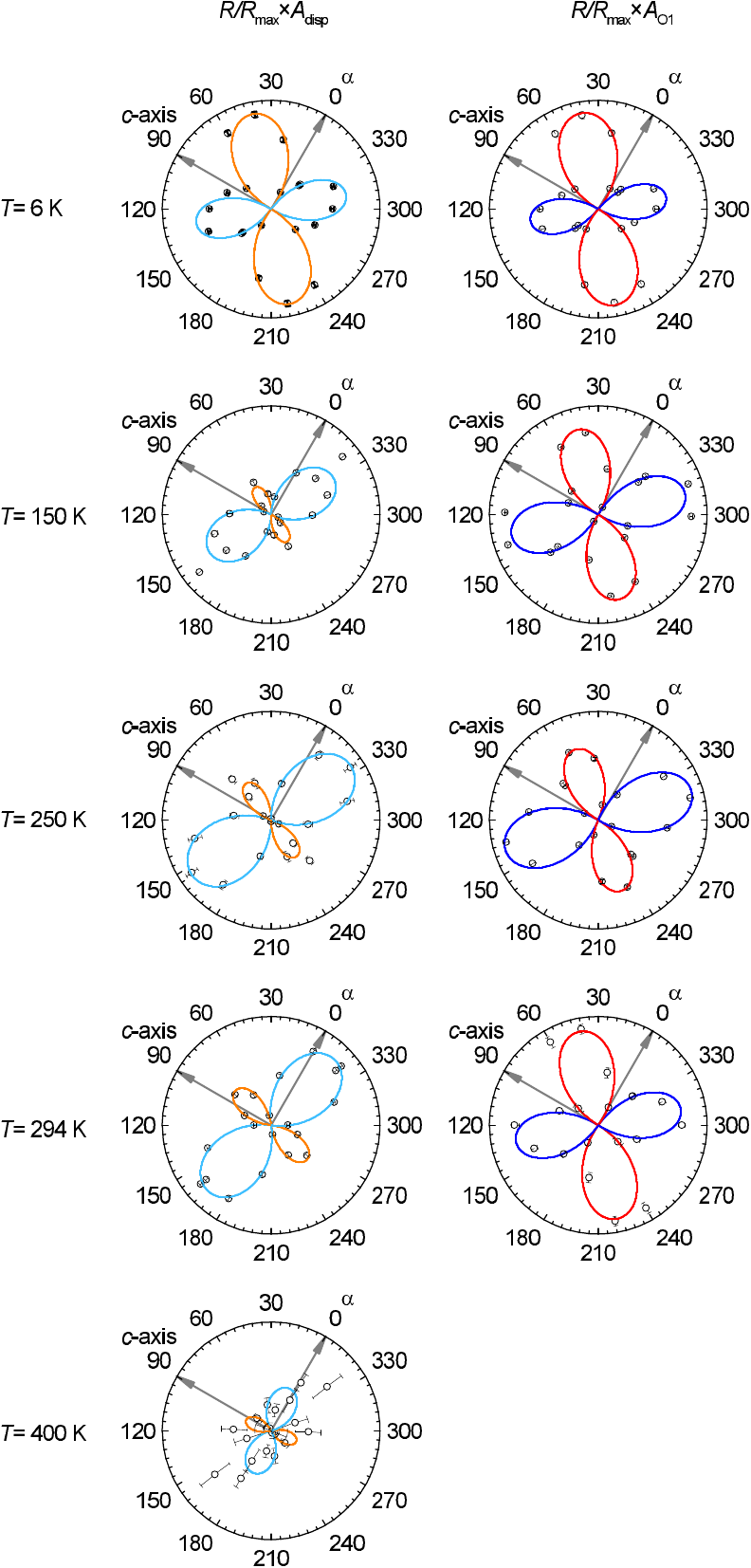} \caption{The main DECP component amplitudes as a function probe polarization
angle at different $T$. The left and right column correspond to the
ERDM and oscillatory mode O1, respectively. The solid lines are fits
using Eq. (\ref{eq:Aangular}) where the red colors correspond to
the positive lobes, and the blue colors to the negative lobes.}
\label{fig:CMpolar} 
\end{figure}

Due to the strong static reflectivity anisotropy it is instrumental
to measure and analyze the $T$-dependence of the unnormalized $\Delta R$
anisotropy. In Fig. \ref{fig:CMpolar} we show the angular dependence
of the amplitude for the two most prominent DECP components at a few
characteristic temperatures. The angular-dependent DECP amplitudes
at any given $T$ were obtained by means of a global fit of Eq. (\ref{eq:fitfunc})
to $\Delta R$ measured at different polarization angles. In the global
fit only the amplitudes were allowed to vary with the polarization
angle whereas the frequencies, dampings and relaxation times were
kept angle independent. Similar as the total $\Delta R_{\mathrm{max}}$
{[}see Fig. \ref{fig:figDRAniso} c){]} the angular dependence of
the individual DECP amplitudes can be well described with a phase
shifted harmonic function with an offset:

\begin{equation}
A(\theta)\propto1+C_{3}\cos[2(\theta-\theta_{0})].\label{eq:Aangular}
\end{equation}
At low $T$ the angular phase shift, $\theta_{0}\sim40^{\circ}$,
is similar for all DECP components. With increasing $T$, however
the oscillatory components show $T$-independent $\theta_{0}\sim40^{\circ}$,
while the ERDM shows a gradual rotation towards $\theta_{0}\sim90^{\circ}$.

\begin{figure}[ht!]
\includegraphics[width=0.9\columnwidth]{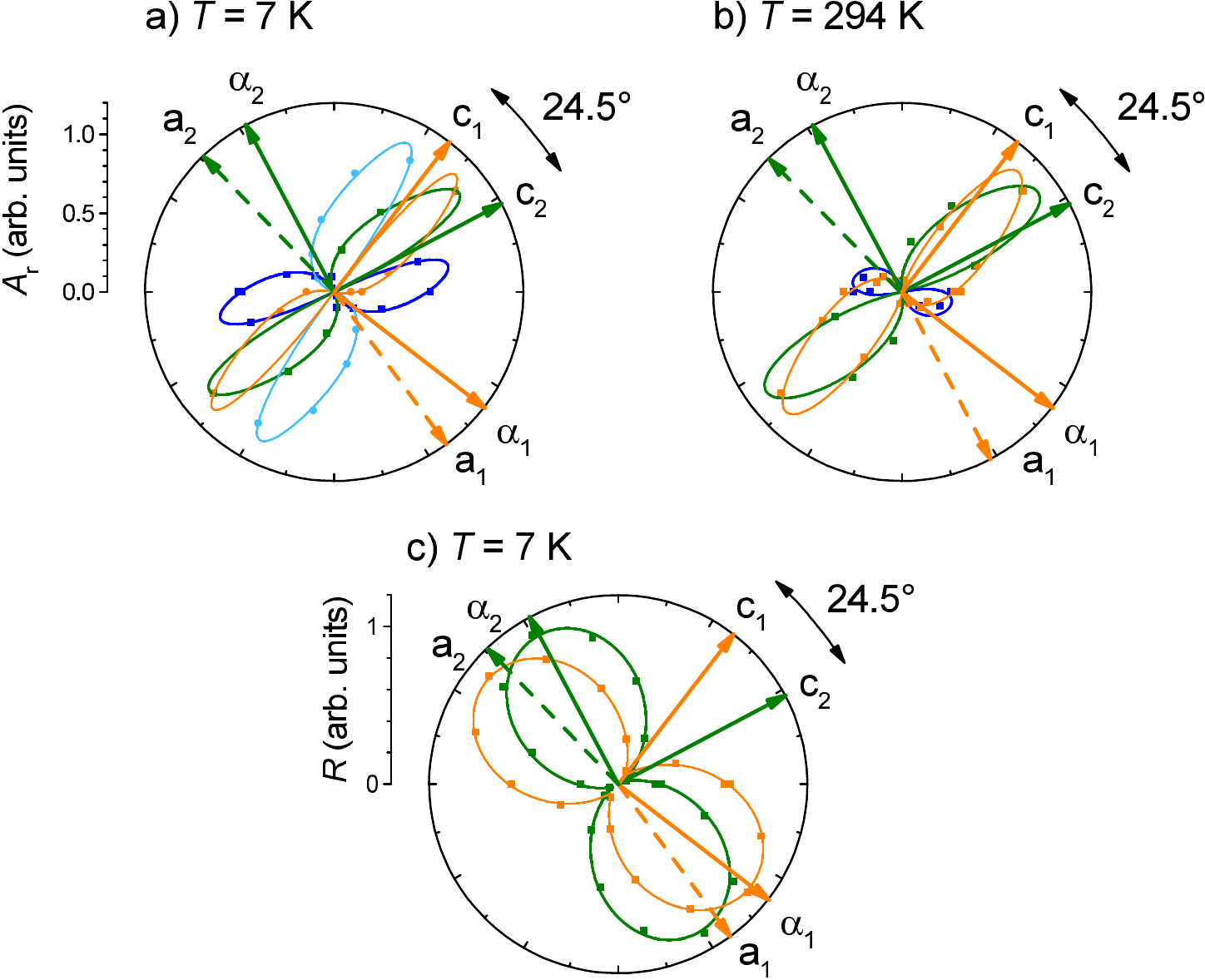} \caption{Comparison of the transient reflectivity anisotropy in growth twin
crystals. (a) and (b) transient reflectivity amplitude at two different temperatures: $T=7$\,K and $T=294$\,K
respectively. (c)
Reflectivity at $T=7$ K. The data are normalized for the sake of
comparison. The orientation of the axes was inferred from the static reflectivity. The blue colors indicate the negative lobes.}
\label{fig:twins} 
\end{figure}

We also found that the transient-reflectivity angular dependence in
growth-mirror-twins is mirrored across the twin mirror plane. The
observed angle between the optically determined $c$-axes of the mirror-twins
is $\sim25^{\circ}$ which is close to $32.5^{\circ}$, the angle
expected for the very common \{100\} monoclinic twins. The difference
can be attributed to the $\sim\pm5^{\circ}$ error of the experimentally
determined reflectivity tensor (Eq.~\ref{eq:r-tensor}) orientation
relative to the crystal axes.

\section{Discussion}

\subsection{Symmetry considerations}

Similar as the Raman tensor the optical transient dielectric tensor
$\Delta\hat{\epsilon}$ can be reduced according to the irreducible
representations of the crystal point group. The high-$T$ point group
($C_{2\mathrm{h}}$) has center of inversion where A$_{\mathrm{g}}$
and B$_{\mathrm{g}}$ -symmetry modes can couple to $\Delta\hat{\epsilon}$.
The B$_{\mathrm{g}}$-symmetry tensor \cite{aroyo2011crystallography}
has zero $ac$-plane components and does not apply to our experimental
geometry, while the $ac$-plane part of the A$_{\mathrm{g}}$-symmetry
tensor has the lowest possible symmetry for a symmetric tensor:

\begin{equation}
\Delta\hat{\epsilon}_{\mathrm{ac}}^{\mathrm{Ag}}=\left(\begin{array}{cc}
\Delta\epsilon_{\alpha\alpha} & \Delta\epsilon_{\mathrm{d}}\\
\Delta\epsilon_{\mathrm{d}} & \Delta\epsilon_{\mathrm{cc}}
\end{array}\right).\label{eq:DepsBg}
\end{equation}
Upon loosing center of inversion in the low-$T$ point group (C$_{s}$),
the A$_{\mathrm{g}}$ and B$_{\mathrm{g}}$ representations of $C_{2h}$
reduce to A' and A'' representations, respectively, without changes
in the allowed Raman tensor components. The A''-symmetry tensor therefore
also does not apply to our experimental geometry so all the observed
modes correspond to the A' representation. In addition, A$\mathrm{_{u}}$
and B$_{\mathrm{u}}$ representations also reduce to A' and A'' representations,
respectively, enabling coupling of the polar modes to $\Delta\hat{\epsilon}$
below $T_{\mathrm{IC}}$. The A' modes have allowed dipole moment
lying in the $ac$-plane.

When measuring the angular dependence of the transient reflectivity
the transient dielectric tensor shape (Eq.~\ref{eq:DepsBg}) would
result in maximum/minimum at an arbitrary angle depending on the relative
values of the three independent components. The gradual rotation of
the ERDM with $T$ can therefore be assigned to different $T$-dependences
of the $\Delta\hat{\epsilon}_{\mathrm{ac}}$ components. The different
$T-$dependencies indicate, however, that the nature of the involved
electronic states changes with $T$. This is further corroborated
with an appearance of the additional picosecond component and increase
of intensities of Raman modes below $175-200$\,K.

\subsection{Electronic relaxation and the non-oscillatory components}

To investigate the single-particle excitations, it is interesting
to consider the data in Fig.~\ref{fig:AvsT} in more detail. As discussed
above, the two non-oscillatory components ERDM and e1, associated
with the overdamped electronic relaxation processes, show an unusual
temperature evolution around $T_{\mathrm{el}}$ where the resistivity
hump occurs. On the other hand, they show no conventional critical
behavior of their relaxation time, that could be expected at the CDW
transitions at $T_{\mathrm{IC}}$ or $T_{\mathrm{IC-C}}$. In contrast,
an increase of the relaxation times, in particular $\tau_{\mathrm{e1}}$
is quite prominent close to $T_{\mathrm{el}}$, which could indicate
a critical point that is not related to the established transitions
at higher temperatures.

A picosecond time scale in electronic relaxation and an increase of
the amplitude of the corresponding electronic response with cooling
is expected in gapped systems. From the electronic structure calculations~
\cite{Whangbo,DOSpaper}, we expect a single-particle density-of-states
characteristic of a semi-metal or a narrow band semiconductor
($2\Delta\sim0.1$\,eV). Indeed, one can understand the $T$-dependencies
of the amplitudes $A_{\mathrm{displ}}$ and $A_{\mathrm{e1}}$ in
terms of a gapped behavior in single-particle excitations. To look
into this quantitatively, we apply a Rothwarf-Taylor type bottleneck
model~ \cite{RothwarfTaylor,KabanovModel}, that describes electronic
relaxation via phonons in systems with a $T$-independent gap of a
magnitude $2\Delta$. Within this model the change of reflectivity
is 
\begin{equation}
\frac{\Delta R}{R}\propto n_{\mathrm{ph}}\propto[1+\gamma\exp(-\frac{\Delta}{k_{B}T})]^{-1},\label{eq:Tindepbottleneck}
\end{equation}
where $n_{\mathrm{ph}}$ corresponds to the photoexcited single particle
excitations density. The parameter $\gamma=\frac{2\nu}{N(0)\hbar\Omega_{C}}$
is the ratio between the number of the involved bosonic and electronic
degrees of freedom. Here $\hbar\Omega_{C}\sim2\Delta$ corresponds
to the characteristic bottleneck-phonon frequency, $\nu$ to the number
of the involved phonon modes and $N(0)$ to the number of the involved
electronic states.

The temperature dependence of $A_{\mathrm{displ}}$ is fitted well
with this formula, as shown in Fig \ref{fig:AvsT} (a). The fit yields
$2\Delta=137\pm10$\,meV and $\gamma\sim60$. Based on the LDA it
is reasonable to take $N(0)\sim10$\,eV$^{-1}$~ \cite{DOSpaper},
which yields $\nu\sim30$. The number is similar to $\nu=10-20$,
obtained in the cuprates \cite{KabanovModel}, \cite{PhysRevB.84.174516}.

We also fit $A_{\mathrm{displ}}$ within a temperature-dependent BCS-type
gap bottleneck model~ \cite{KabanovModel} $\Delta_{\mathrm{BCS}}(T)$
with $T_{\mathrm{c}}=T_{\mathrm{IC}}\sim350$\,K to see whether one
could interpret the results in terms of the opening of a CDW gap.\footnote{To fit the finite $A_{\mathrm{displ}}$ above $T_{\mathrm{IC}}$ in
the case of $T$-dependent gap we assume a small $T$-independent
background.} In this case 
\begin{eqnarray}
\frac{\Delta R}{R} & \propto & [\Delta_{\mathrm{BCS}}(T)+\frac{k_{B}T}{2}]^{-1}\nonumber \\
 &  & \times\{1+\gamma\sqrt{\frac{k_{B}T}{\Delta_{\mathrm{BCS}}(T)}}\exp[-\frac{\Delta_{\mathrm{BCS}}(T)}{k_{B}T}]\}^{-1},\label{eq:Tdepbottleneck}
\end{eqnarray}
yields similar parameter values: $2\Delta_{\mathrm{BCS}}(0)=112\pm10$\,meV
and $\gamma\sim45$. The BCS type temperature dependent gap function
describes the amplitude, $A_{\mathrm{_{disp}}}$, equally well as
the temperature independent one, so the temperature dependence of
the gap at high temperatures cannot be discerned from these fits alone.

On the other hand, the absence of a relaxation time divergence at
$T_{\mathrm{IC}}$ and $T_{\mathrm{IC-C}}$ strongly suggests that
there is no rapid BCS-like change in the DOS at the IC and IC-C transitions,
so a T-independent gap is more appropriate in this case.

Within the same approach, we also considered the temperature dependence
of the picosecond component amplitude, $A_{\mathrm{e}1}$. In order
to analyze this we subtracted from $A_{e1}$ a part proportional to
$A_{\mathrm{displ}}$ to remove the negative high-temperature tail
shown in Fig. \ref{fig:AvsT} a) \footnote{It is conceivable that
the displacive component can not be completely described by an exponential
fit and the displacive component residual is picked by component e1.}  Both the temperature
independent and the BCS-type model again can fit the
temperature dependence of $A_{e1}$ quite well. However, the temperature-independent-gap
fit yields an unphysical value of $\gamma\sim10000$, which suggests that a temperature
independent gap is not appropriate in this case. Rather, assuming
a temperature-dependent BCS-type gap yields $2\Delta_{\mathrm{BCS}}(0)=85\pm10$\,meV,
$\gamma\sim26$ and $T_{\mathrm{c1}}=160\pm5$ K. Importantly, the
peak in $\tau_{\mathrm{e}1}$ observed just below $T_{\mathrm{c1}}$
is consistent with opening of an additional low-$T$ gap.

The relatively decoupled transient reflectivity dynamics of the two
exponential components despite similar gap magnitudes suggests that
the high-$T$ and low-$T$ gaps are separated in $k$-space and the
momentum scattering (impurity and acoustic phonon) between the relevant
electronic states must be weak. These data suggest that the band structure contains multiple gaps, in different regions
of the Brillouin zone.

\subsection{Oscillating modes}

We now turn to a closer discussion of the oscillatory modes. The dominant
observed coherent mode O1 appears below $T=360$\,K and shows an
unusual behavior of the frequency and damping. It directly corresponds
to the lowest frequency Raman mode (RO1). In Raman spectra (RS) the
mode does not show a significant intensity change across the IC and
IC-C transitions so it is not a folded mode or the IC-transition soft
mode. The weak $T$-dependence of the mode RS intensity indicates
that the main contribution to the strong O1-mode coherent amplitude
$T$-dependence is due to the $T$-dependence of the displacive component
drive amplitude. The minor difference
between the $T-$dependence of $A_{\mathrm{O}1}$ and $A_{\mathrm{disp}}$
can be attributed to the temperature dependence of $\tau_{\mathrm{disp}}$\cite{zeiger1992theory}
and/or slight $T$-dependence of the mode coupling to the reflectivity.
Moreover, component e1 in principle also contributes to the displacive
drive. This is not explicitly included in the model (\ref{eq:fitfunc}),
but, since $A_{\mathrm{O}1}$ and $A_{\mathrm{disp}}$ are independent
fit parameters $A_{\mathrm{O}1}$ would pick such contribution. 

The strong softening of $\sim0.2$\,THz observed in a relatively
narrow $T$ range around the nominal $T_{\mathrm{IC}}\sim350$\,K
is absent in RS. This indicates that the coherent oscillations observed
at $T=360$ K most likely correspond to weaker mode O3 that is obscured
by mode O1 in the 350-250\,K $T$-range and is detectable only at
lower temperatures. Both O1 and RO1 show strong softening upon cooling
to $T_{\mathrm{m}}\sim200$\,K and then harden back with
further cooling. The RO1 minimum frequency appears slightly lower
than the O1 minimum frequency.

The $T$-dependence of $\Omega_{\mathrm{O1}}$ is anomalous and clearly
unrelated to the ordering either at $T_{\mathrm{IC}}$ of $T_{\mathrm{IC-C}}$.
The mode-O1 maximum damping is near $T_{\mathrm{m}}$,
rather than near the either of the transition temperatures\footnote{Similar tendency is also found in the Raman mode width, albeit
proximity to the notch-filter cutoff prevents reliable linewidth temperature
determination}. Since it is not a folded mode it cannot be explained within the
simple CDW-like folded-phonon-mode scenario \cite{schaefer2014collective}
related to the IC transition. Its behavior would be the most similar
to the folded CDW mode behavior in the case of a strong (``non-adiabatic'')
coupling between a phonon and a CDW electronic mode (EM). \cite{schaefer2014collective}
When the temperature dependent relaxation rate of the EM crosses the
phonon frequency, their mixing causes a phonon softening anomaly and
strong critical broadening with increasing temperature. But in our
case no critical EM is observed and the broadening near $T_{\mathrm{IC}}$
is much smaller than in the case of a CDW mode.

The other coherent modes O2-O4, and the remaining weak modes (that
were not fit), obtain a considerable intensity well below $T_{\mathrm{IC}}$
an $T_{\mathrm{IC-C}}$ but show a much weaker frequency $T$-dependence.
The modes are not completely resolved in RS due to the worse frequency
resolution and appear as weaker features in RS. Similar to mode RO1
they are present\footnote{Except the 2.6-THz mode that is rather weak in both, RS and the coherent
response.} above $T_{\mathrm{IC}}$ in RS and show no abrupt intensity/shift
change at $T_{\mathrm{IC}}$ and $T_{\mathrm{IC-C}}$. The strong
$T$-dependences of their amplitudes in the coherent response are
therefore attributed as for mode O1 to the $T$-dependence of the
displacive-component-drive amplitude.

The coherent modes O2-O4 frequency $T$-dependencies are small and
resemble the shape of the $\Omega_{\mathrm{O1}}$ vs $T$ dependence
below $\sim150$ K, presumably due to repulsion resulting from their
weak coupling\footnote{The coupling has to be weak since no significant broadening of the
modes is observed despite significant overlap with mode O1 at low
$T$.} to mode O1. Their smaller frequency shift indicates that they do
not play an active role in the electronic changes observed below $\sim200$\,K.

In addition to the coherent response RS also show sensitivity to the
electronic changes observed below $\sim200$\,K. Some modes
show a clear increase of intensity. There are also some additional
frequency shifts below $175-200$\,K that are the most notable
for the 617\,cm$^{-1}$ and\,375\,cm$^{-1}$ modes.

The IC and IC-C transitions appear more clearly in the behavior of
some higher frequency Raman modes. These modes show small frequency
shifts (either hardening or softening) that can be well correlated
with the IC/C structural order parameter shown in Fig. \ref{fig:CM}
a). (see for example the 525\,cm$^{-1}$ and 748\,cm$^{-1}$ modes
in Fig. \ref{fig:Raman}). Some of these modes show also additional
frequency anomalies below $\sim200$\,K (see for example
the 391\,cm$^{-1}$ and 639\,cm$^{-1}$ modes).

\subsection{The enigmatic state at low $T$}

Based on the discussion above, the transient reflectivity across the
IC and IC-C transitions of \mox does $not$ behave as a typical CDW
system. The anomalous behavior in \mox at low $T$ is apparently
driven by something seemingly quite unrelated to either of the 350\,K
(IC) or 285\,K (C) transitions. It is particularly interesting that
the electronic relaxation shows clear evidence of a gap opening upon cooling below
$T\sim150$\,K, even showing signs of a divergence of the lifetime
near this temperature, while at the same time, the resistivity starts to drop. A simple explanation is that the carriers that dominate the
transport properties are separated in $k$-space from the gapped excitations
that are observed in the time-resolved spectroscopy.
However, the anomalies in both channels seemingly capture the
temperature evolution of the electronic structure beyond the reported
transitions.

Although we observe a gap-related bottleneck below $T_{\mathrm{IC}}$,
our data suggest that the low-energy electronic band structure evolves
rather smoothly across the IC(-C) transitions without a large immediate
DOS loss at $T_{\mathrm{IC}}$ and $T_{\mathrm{IC-C}}$. The anomalous
large softening of mode O1 below 350\,K and the appearance of the
additional Raman mode below $175-200$\,K might be explained
by the existence of an additional hidden electronic ordering transition
at low temperatures.

On the other hand, the IC/C structural order parameter \cite{Sato}
(in Fig. \ref{fig:CM} a) shows a large continuous increase below
$\sim280$\,K down to $\sim100$\,K. The continuously increasing
structural order parameter together with a possible chemical potential
shift could lead to a Lifshitz transition \cite{ZHANG2017950,Chi2017}
and/or an additional gap opening below $175-200$ K. The
observed low-$T$ gap opening in the transient response and changes
in transport \cite{DOSpaper} would be compatible also with such scenario.
However, the large mode O1/RO1 softening and the changes in
the Raman spectra below $175-200$\,K cannot be directly linked
to the structural order parameter induced changes. The mode O1/RO1
softening could be associated with an incipient Fermi-surface driven
instability that is eventually suppressed by the competing IC/C-order-induced
low-energy DOS suppression while the absence of the 617\,cm$^{^{-1}}$
above $175-200$\,K would be attributed to a stronger decrease
of the resonant Raman cross-section in comparison to the other modes. 

The incipient transition scenario is also favored by the discrepancy
of the maximum softening temperature, $T_{\mathrm{m}}\sim200$\,K,
that is somewhat higher than the low-$T$ gap opening temperature
$T_{\mathrm{c1}}\sim160\:\mathrm{K\sim}T_{\mathrm{el}}$. 

Another possible explanation for the difference
between $T_{c1}$ and $T_{\mathrm{m}}$ would be the presence of a rather broad fluctuation region where different couplings
of different probes result in the apparently different transition temperatures as was observed for the pseudogap in the cuprates.

\section{Summary and Conclusions}

The transient-reflectivity and vibrational Raman spectra in \mox
single crystals reveal significant details on the evolution of the
electronic structure and collective mode behavior associated with
different ordering phenomena in this unusual material.

Below the incommensurate transition at $T\sim350$\,K the temperature
evolution of the fastest sub-picosecond transient relaxation component
indicates a bottleneck due to the presence of a temperature independent
(pseudo-~) gap with a magnitude $2\Delta=140$\,meV. The absence
of any significant singularity in any of the transient reflectivity
components at either the incommensurate transition temperature $T_{\mathrm{IC}}\sim350$
K, or the incommensurate-commensurate transition at $T_{\mathrm{IC-C}}=285$\,K
suggests only a minor modification of the low-energy DOS takes place
at these transitions. Moreover, no additional Raman active vibrational
modes appear at either of the transitions that could be assigned to
a soft mode, so the behavior at the IC and IC-C transitions is highly
unconventional.

Even more enigmatic is the picosecond relaxation component dynamics
that suggests an opening of a previously unknown temperature-dependent
gap $2\Delta(0)\sim90$\,meV around $T=160$\,K, that is apparently
unrelated to the previously observed transitions. \cite{Sato} The
most prominent coherent mode and its Raman counterpart at $\mbox{\ensuremath{\sim}}2$\,THz
show anomalous $\sim20$ \% softening around $T_{\mathrm{m}}\sim200$
K, with accompanying anomalies of some of the higher frequency Raman
modes. Altogether, the observed anomalous behavior points towards
the presence of a previously unreported hidden electronic transition below $T_{\mathrm{c1}}\sim160$
K.

The Brillouin-zone-center
lattice dynamics anomalies observed well above $T_{\mathrm{c1}}\sim160$
K suggest either a broad fluctuation region associated with the transition
at $T_{\mathrm{c1}}$ or another incipient transition that is eventually
suppressed by the hidden electronic order. Further structure-sensitive
experiments focusing towards small wavevectors could possibly shine
more light on this issue.
\begin{acknowledgments}
The authors acknowledge the financial support of Slovenian Research
Agency (research core funding No-P1-0040) and European Research Council
Advanced Grant TRAJECTORY (GA 320602) for financial support, valuable
discussions with V.V. Kabanov, Y.A. Gerasimenko and I.V. Madan and A. Meden for the initial characterization of molybdenum suboxide phases. JM acknowledges support by Program P1-0044 of Slovenian Research Agency. 
\end{acknowledgments}

\section{Appendix}

\begin{figure}[ht!]
\includegraphics[width=0.9\columnwidth]{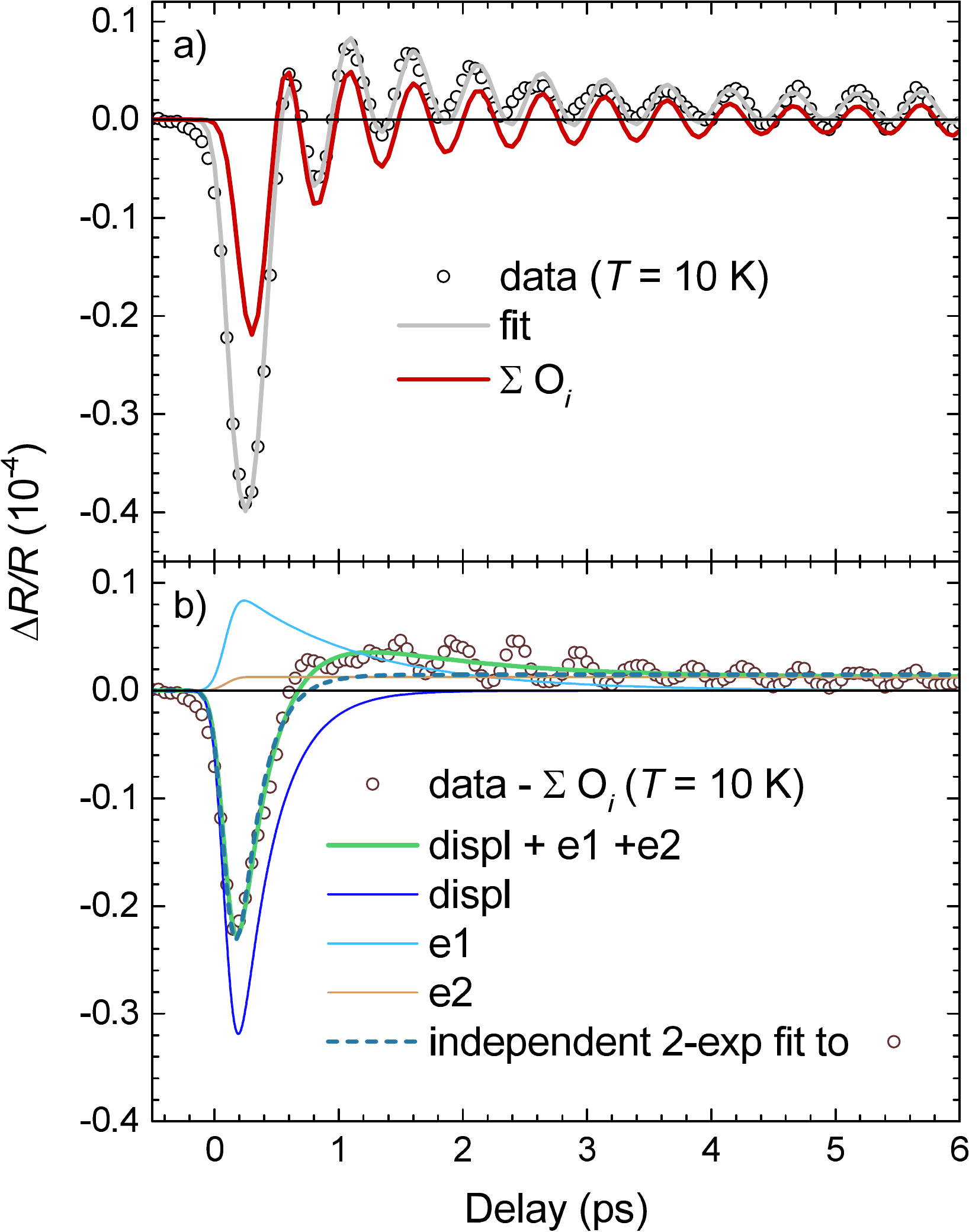} \caption{Decomposition of the transient reflectivity at 10\,K into components.
a) Comparison of the raw transient reflectivity at 10\,K, the fit
(Eq.~\ref{eq:fitfunc}) and the sum of the four fit oscillatory components,~O1-O4.
b) Comparison of the residual of the raw data (with O1-O4 components
subtracted) to the exponential fit components. It is evident that
the sum of components displ, e1 and e2 fits the residual better than
a simple two exponent fit.\label{fig:figDecomposition}}
\end{figure}

In Fig. \ref{fig:figDecomposition} we show decomposition of the full
DECP fit (Eq.~\ref{eq:fitfunc}) into different components. The sum
of the fitted oscillator components at $T=10$ K consisting from oscillators
O1-O4 is shown in Fig. \ref{fig:figDecomposition} a) in comparison
to the total transient reflectivity. Note that the displacive
driven oscillator coordinates contain both, an oscillating and an
exponential part.\cite{zeiger1992theory} The latter is proportional
to the exponential displacive drive and relaxes with the time constant
$\tau_{\mathrm{disp}}$.

The exponential components with the remaining oscillators are clearly
revealed after we subtract the four fitted oscillator responses {[}see
Fig. \ref{fig:figDecomposition} b){]}. The non-oscillating part of
the residue can not be completely fit with a two component exponential
relaxation\footnote{In order to fit the sub-picosecond relaxation and the long-delay offset
at least two exponential components are necessary. One with $\tau\sim\mbox{\ensuremath{\tau}}_{\mathrm{disp}}$
and one with $\tau\rightarrow\infty$.} indicating the presence of picosecond component e1.

%\begin{thebibliography}{54}
 
%\end{thebibliography}

\bibliographystyle{apsrev4-1}
\bibliography{biblioMo8O23prb}

\end{document}